\documentclass[aps,pra,twocolumn,groupedaddress,showpacs]{revtex4-1}

\usepackage{graphicx}
\usepackage{epsfig}
\usepackage{amsmath}
\usepackage{hyperref}
\usepackage{longtable}

\begin{document}

\DeclareGraphicsExtensions{.eps,.EPS,.jpg,.bmp}


\title{Quasi-forbidden 2-body F\"orster resonances in cold Cs Rydberg gas}


\author{B. Pelle}
\affiliation{Laboratoire Aim\'e Cotton, UPR 3321 CNRS, Universit\'e Paris-Sud, ENS Cachan, B\^atiment 505, Campus d'Orsay, 91405 Orsay Cedex, FRANCE}

\author{R. Faoro}
\altaffiliation{Dipartimento di Fisica, Universit\`a di Pisa, Largo Pontecorvo 3, I-56127 Pisa, Italy}
\affiliation{Laboratoire Aim\'e Cotton, UPR 3321 CNRS, Universit\'e Paris-Sud, ENS Cachan, B\^atiment 505, Campus d'Orsay, 91405 Orsay Cedex, FRANCE}

\author{J. Billy}
\altaffiliation[Permanent address:~]{Laboratoire Collisions Agr\'egats R\'eactivit\'e, IRSAMC, Universit\'e de Toulouse-UPS and CNRS, UMR 5589, F-31062 Toulouse, France}

\author{E. Arimondo}
\altaffiliation{Dipartimento di Fisica, Universit\`a di Pisa, Largo Pontecorvo 3, I-56127 Pisa, Italy}

\author{P. Pillet}
\affiliation{Laboratoire Aim\'e Cotton, UPR 3321 CNRS, Universit\'e Paris-Sud, ENS Cachan, B\^atiment 505, Campus d'Orsay, 91405 Orsay Cedex, FRANCE}

\author{P. Cheinet}
\email[]{patrick.cheinet@u-psud.fr}
\affiliation{Laboratoire Aim\'e Cotton, UPR 3321 CNRS, Universit\'e Paris-Sud, ENS Cachan, B\^atiment 505, Campus d'Orsay, 91405 Orsay Cedex, FRANCE}



\date{\today}

\begin{abstract}
Cold Rydberg atoms are known to display dipole-dipole interaction allowed resonances, also called F\"orster resonances, which lead to an efficient energy transfer when the proper electric field is used. This electric field also enables resonances which do not respect the dipole-dipole selection rules under zero field. A few of these quasi-forbidden resonances have been observed but they are often overlooked. Here we show that in cold $^{133}$Cs atoms there is a large number of these resonances that display a significant transfer efficiency due to their strong interactions, even at low electric field. We also develop a graphical method enabling to find all possible resonances simultaneously. The resulting dramatic increase in the total number of addressable resonant energy transfers at different electric fields could have implications in the search for few-body interactions or macro-molecules built from Rydberg atoms.

\end{abstract}


\maketitle



Resonant energy transfer in Rydberg atoms is well known to enhance 2-body interactions between adjacent atoms \cite{1981PRLGallagher}, and is also called F\"orster resonances in analogy with the biological process known as FRET (Fluorescence Resonance Energy Transfer). Those kind of resonances have been studied to a great extent in Rydberg atoms originally in \cite{1981PRLGallagher, 1982PRAGallagher} and for different species in high electric fields \cite{1983PRAGallagher, 1994PRAGallagher}. They continue to be of interest \cite{2004PRLNoel, 2008PRLvanLinden, 2010PRLEntin, 2012PRLPfau, 2014NatPhysicsBrowaeys, 2015PRLBrowaeys, 2015PRABrowaeys} in the study of dipole blockade \cite{2006PRLPillet, 2009NatureGrangier, 2010JOSABPillet, 2010PRAEntin, 2014PRLBrowaeys} and its induced entanglement \cite{2010PRLBrowaeys, 2010NJPBrowaeys}, but also in studies on many body \cite{2004PRAGallagher} or more recently few-body \cite{2012PRLPillet, 2015NatureCommPillet} interactions, or even in an atom interferometer \cite{2012PRXPfau}. The dipole-dipole interactions between Rydberg atoms are also the basis of proposed approaches to realize quantum gates \cite{2000PRLLukin, 2001PRLLukin, 2010PRLSaffman, 2014PRABrowaeys} or for the formation of novel long-range molecules \cite{2000PRLSadeghpour, 2009NaturePfau, 2011SciencePfau, 2012PRAGallagher}.\\

While dipole allowed F\"orster resonances have been studied in many different species, only a few quasi-forbidden resonances have been observed in sodium and potassium \cite{1983PRAGallagher, 1994PRAGallagher}. These resonances exist in presence of an external static electric field leading to some coupling between Rydberg states in the transition dipole matrix, and where the eigenstates of the dipole-dipole Hamiltonian are no longer the pure eigenstates of the system. Then the dipole-dipole selection rules apply only partially, allowing the presence of several quasi-forbidden 2-body F\"orster resonances in addition to the dipole allowed 2-body F\"orster resonance, if one dipole allowed F\"orster resonance exists in a specific atom. For instance, one specific resonance has been used in cesium to observe a 4-body interaction \cite{2012PRLPillet}. These additional resonances could be of interest to increase the number of addressable resonances for possible quantum gates, for the creation of molecules from Rydberg atoms or for all applications based on few-body interactions. Moreover these resonances, if not properly taken into account, could modify the results of the previously described processes based on the dipole allowed resonances. In addition, these resonances could introduce addressable resonances whenever no dipole allowed resonance exists, for instance in cesium for $n > 42$ or when starting from an initial $ns$ state.\\

Within the present work, using cold $^{133}$Cs atoms excited to Rydberg states, we show the presence of a large number of quasi-forbidden resonant energy transfers at low electric field, where we can identify each resonance and compute their dipole-dipole couplings. To determine their resonant Stark field, we apply a graphical method which is well suited to predict resonances in the vicinity of the multiplicity and which is based on a combination of two Stark diagrams to solve the resonance condition.

\section{Dipole-dipole interactions}
\label{sec:Dip-dip}

\begin{figure}[h]
    \begin{center}
    	\includegraphics[width=7.5 cm]{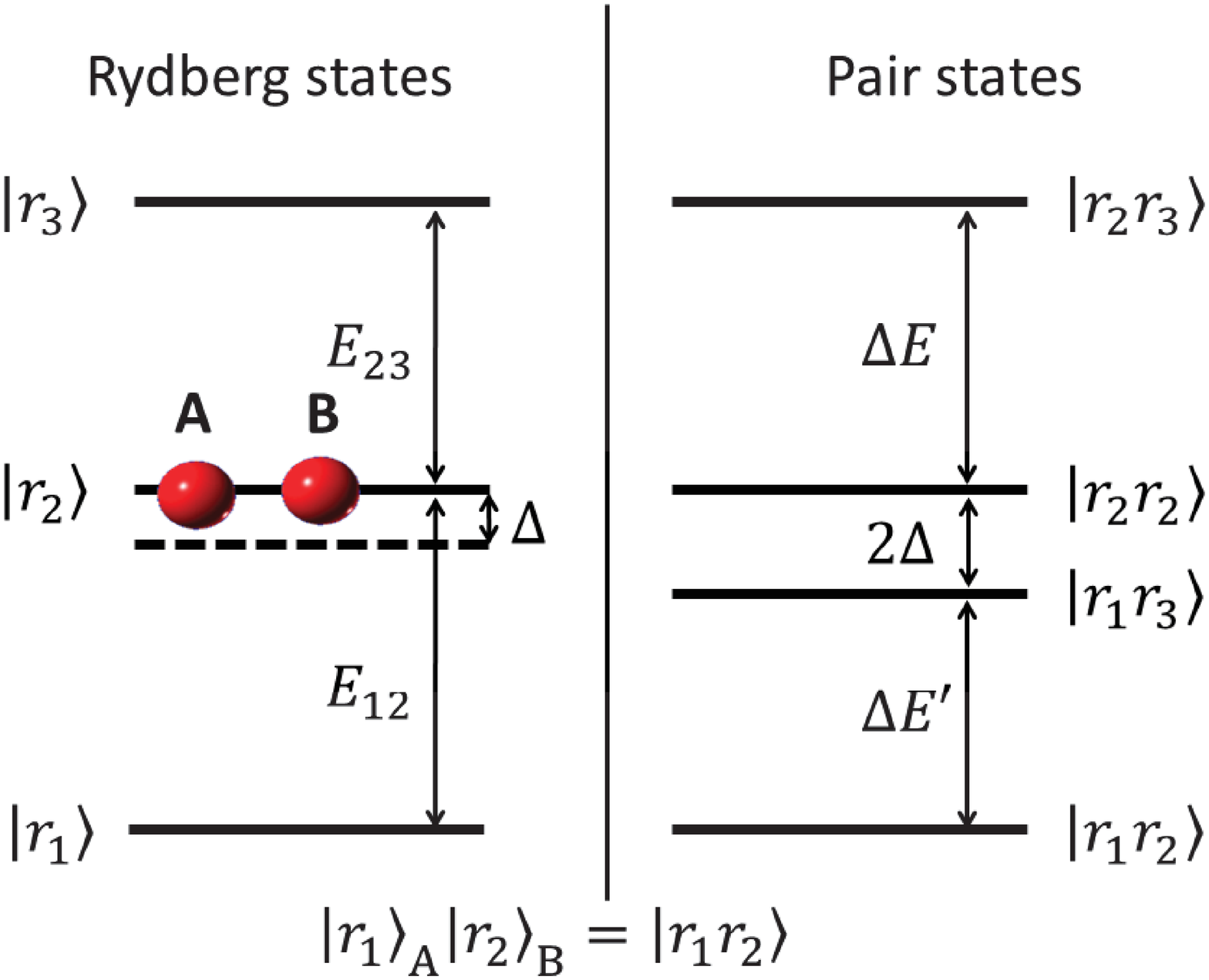}
		\end{center}
   	\caption{(Color online) Single atom basis (left) and two-atom basis (right) for the description of interactions between atoms $\mathrm{A}$ and $\mathrm{B}$ having three Rydberg states ${\left|r_{1}\right\rangle}$, ${\left|r_{2}\right\rangle}$, and ${\left|r_{3}\right\rangle}$.}
    \label{Fig:TwoAtomsBasis}
\end{figure}

We consider two atoms $\mathrm{A}$ and $\mathrm{B}$, separated by a distance ${R \boldsymbol{n}}$ and originally both in a given Rydberg state ${\left|r_{2}\right\rangle}$. We also take into account two other Rydberg states ${\left|r_{1}\right\rangle}$ and ${\left|r_{3}\right\rangle}$, almost equally separated from the initial state. We then define the two-atom state basis as ${\left|r_{1} r_{3}\right\rangle = 1/\sqrt{2} \left(\left|r_{1}\right\rangle_{\mathrm{A}} \left|r_{3}\right\rangle_{\mathrm{B}} + \left|r_{3}\right\rangle_{\mathrm{A}} \left|r_{1}\right\rangle_{\mathrm{B}}\right)}$, sketched in Fig.~\ref{Fig:TwoAtomsBasis}. The dipole-dipole interaction can couple the starting two-atom state ${\left|r_{2} r_{2}\right\rangle}$ with other states, for instance ${\left|r_{1} r_{3}\right\rangle}$, and is then described by~\cite{1994BookGallagher}:
\begin{eqnarray}
\hat{V}_{\mathrm{dd}} & = & \left\langle r_{2} r_{2} \right| \hat{H}_{\mathrm{dd}} \left| r_{1} r_{3}\right\rangle \nonumber\\
& = & \frac{1}{4 \pi \varepsilon_0} \frac{ \boldsymbol{\hat{\mu}}_{r_{1} r_{2}} \cdot \boldsymbol{\hat{\mu}}_{r_{2} r_{3}} - 3 (\boldsymbol{\hat{\mu}}_{r_{1} r_{2}} \cdot \boldsymbol{n})(\boldsymbol{\hat{\mu}}_{r_{2} r_{3}} \cdot \boldsymbol{n})}{R^3}
\label{eq:DipoleDipoleCoupling}
\end{eqnarray}
where $\hat{V}_{\mathrm{dd}}$ is the dipole-dipole interaction, $\hat{H}_{\mathrm{dd}}$ the associated Hamiltonian, $\varepsilon_0$ the electric vacuum permittivity, $\boldsymbol{\hat{\mu}}_{r_{1} r_{2}} = \left\langle r_{1}|\boldsymbol{\hat{\mu}}| r_{2}\right\rangle$ and $\boldsymbol{\hat{\mu}}_{r_{2} r_{3}} = \left\langle r_{2}|\boldsymbol{\hat{\mu}}| r_{3}\right\rangle$ the transition dipole matrices between the involved atomic states with $\boldsymbol{\hat{\mu}}$ the electric dipole moment operator. Due to selection rules on the angular part of the zero-field Rydberg wave functions and on parity conservation, the dipole moment operator introduces the selection rules ${\Delta l = \pm 1}$ and ${\Delta m_{j}=0, \pm 1}$ in the zero-field Rydberg states basis, with $l$ the orbital quantum number and $m_{j}$ the second total angular momentum quantum number.\\


In general, ${\left|r_{2} r_{2}\right\rangle}$ and ${\left|r_{1} r_{3}\right\rangle}$ present an energy mismatch of $2\Delta$ and the dipole-dipole interaction turns into an effective van der Walls interaction. The energy mismatch $\Delta$, also called F\"orster defect, is defined by ${\Delta = E_{12}-E_{23}}$ with $E_{ij}$ the energy difference between the involved Rydberg levels (see Fig.~\ref{Fig:TwoAtomsBasis}). In the rare degenerate case or when an electric field is used to Stark shift the levels into resonance, a strong dipole-dipole interaction is restored between the two atoms and a resonant energy transfer occurs from ${\left|r_{2} r_{2}\right\rangle}$ to ${\left|r_{1} r_{3}\right\rangle}$. It can also be seen as a mutual exchange of excitation between the two atoms. The resonance condition is then expressed as follows, with $E_{i}$ the energy of the state $i$:
\begin{equation}
( E_{r_{3}} - E_{r_{2}} ) - ( E_{r_{2}} - E_{r_{1}} ) = 0 \Leftrightarrow \Delta = 0.
\label{eq:ResonanceCondition}
\end{equation}

\begin{figure}[h]
    \begin{center}
    	\includegraphics[width=5.5 cm]{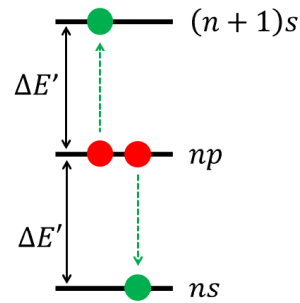}
		\end{center}
   	\caption{(Color online) Energy levels involved in the allowed 2-body F\"orster resonance, ${2 \times n p \leftrightarrow ns + (n+1) s}$, with its population transfer during the F\"orster resonance from the initial states (in red) to the final states (in green).}
    \label{Fig:2bForsterResonance}
\end{figure}

A well known set of allowed 2-body Stark-tuned F\"orster resonances in $^{133}$Cs is between $np$ and $ns$ states, as shown on the Fig.~\ref{Fig:2bForsterResonance} and expressed as:
\begin{equation}
2 \times n p \leftrightarrow ns + (n+1) s
\label{eq:ForsterResonance}
\end{equation}
where $n$ is the principal quantum number and $s$, $p$ denote the orbital quantum numbers. Those allowed resonances give for instance a dipole-dipole coupling of $112$~MHz at 1~$\mu$m for the process of Eq.~(\ref{eq:ForsterResonance}) with ${28p_{3/2}m_{j}\!=\!3/2}$ as the initial state, and $210$~MHz at 1~$\mu$m with ${32p_{3/2}m_{j}\!=\!3/2}$.\\

In presence of a static electric field, the orbital quantum number $l$ is no longer a good quantum number even if we will still use it as a convenient and unique way to label states. With a classical field $\boldsymbol{F}$, each pair of states ${\left|r_{1}\right\rangle}$ and ${\left|r_{2}\right\rangle}$ satisfying ${\Delta l = \pm 1}$ is coupled due to the Stark effect~\cite{1979PRAZimmerman}:
\begin{equation}
\left\langle r_{1} \right| \hat{H}_{\mathrm{F}} \left| r_{2}\right\rangle = - \boldsymbol{\hat{\mu}}_{r_{1} r_{2}} \cdot \boldsymbol{F}
\label{eq:StarkCoupling}
\end{equation}
with $\hat{H}_{\mathrm{F}}$ the Stark Hamiltonian. The new eigenstates then contain a combination of $l$ states instead of pure states. Then the selection rule on the orbital quantum number is relaxed and transitions for any $\Delta l$ can be addressed with a strength depending on the $l$-mixing coupling. As an example to compare with the allowed F\"orster resonances, a quasi-forbidden resonance like 
\begin{equation*}
2 \times 28 p_{3/2} m_{j}\!=\!3/2 \leftrightarrow 28 s_{1/2} m_{j}\!=\!1/2 + 25 f_{7/2} m_{j}\!=\!1/2
\label{eq:ForbiddenResExample}
\end{equation*}
reaches a dipole-dipole coupling of $6.65$~MHz at 1~$\mu$m.\\



\begin{figure*}[h]
    \begin{center}
    	\includegraphics[width=17.8 cm]{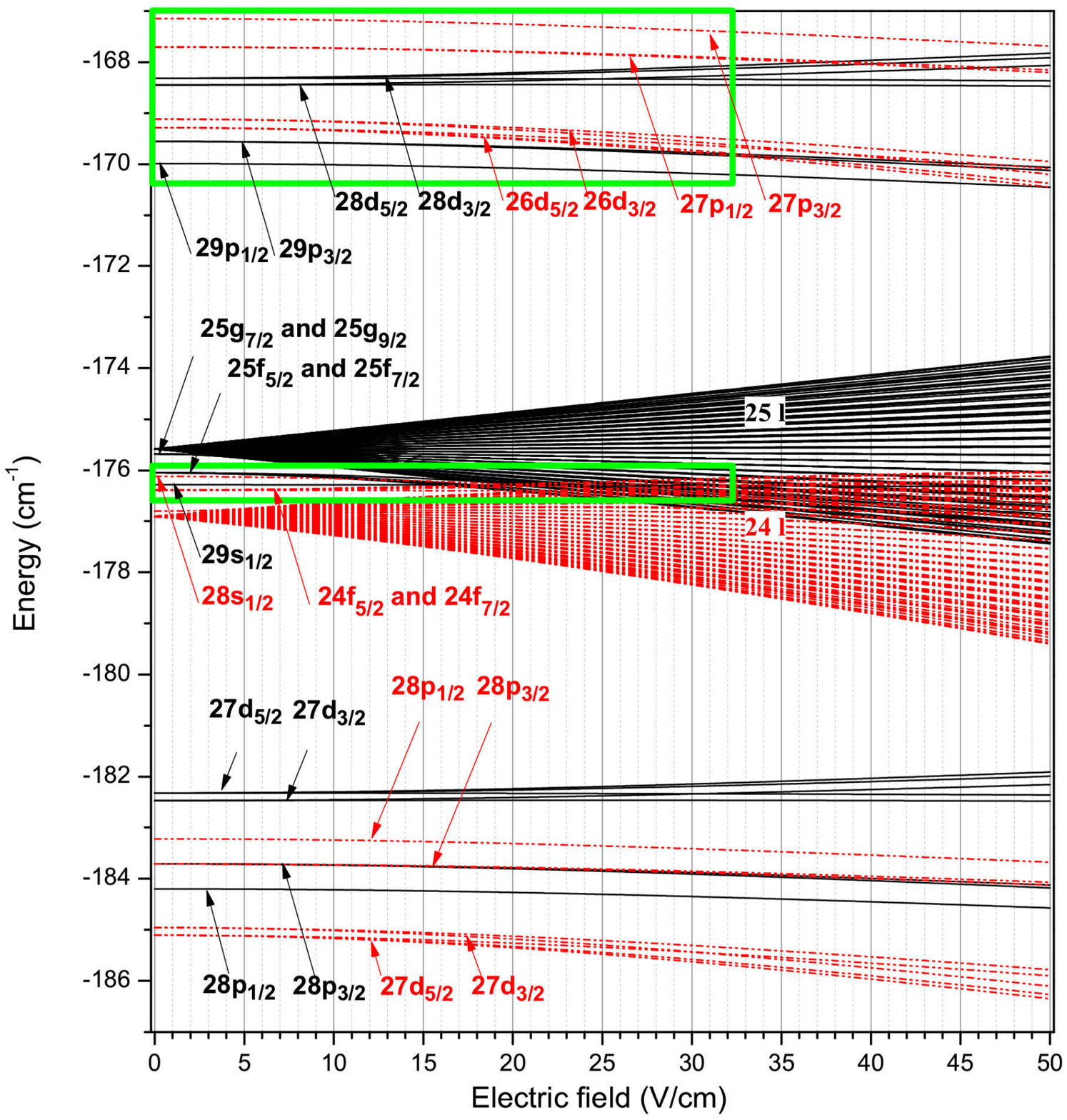}
		\end{center}
		\setlength\abovecaptionskip{-3 cm}
   	\caption{(Color online) Resonance map for the $\left| r_2 \right> = \left| 28 p_{3/2} m_{j}\!=\!3/2 \right>$ initial state. In solid black are plotted the eigenenergies $E_{r_1}$ \textit{vs} the applied external electric field $F$. In dashed red, we plot the energy corresponding to ${2 \times E_{r_2} - E_{r_3}}$ to obtain a graphical solution of the resonance condition occurring in a F\"orster resonance. Only the absolute values of $|m_{j}|$ are plotted. We emphasize the fact that the $\left| r_2 \right>$ initial state curves in solid black and in dashed red are superimposed. In green, we explicit the areas represented in Fig.~\ref{Fig:28p32m32ThForsterResonances28s-25f28s-25g} and Fig.~\ref{Fig:28p32m32ThForsterResonances29p-26d}.}
    \label{Fig:28p32m32ThGlobal}
\end{figure*}

The eigenstate energies $E_{i}$ can be computed as a function of the applied electric field $\boldsymbol{F}$ in a so-called Stark map, as shown in solid black in Fig.~\ref{Fig:28p32m32ThGlobal} in the upper vicinity of the ${28p_{3/2} m_{j}\!=\!3/2}$ state and where we only plot the absolute values $|m_{j}|$. The $l \leq 5$ states in cesium are not degenerated with the other states because of the so-called quantum defect~\cite{1979PRAZimmerman, 1994BookGallagher} originating from the interaction between the Rydberg electron and the ionic core. Then at low values of the electric field, those Rydberg states experience a quadratic Stark effect. On the contrary, states with higher $l$ are degenerated in the multiplicity at low electric field and show a strong linear Stark effect. As we will use this Stark map to identify our quasi-forbidden resonances, our numerical solution of the time independent Schr\"odinger equation takes into account only the second total angular momentum quantum numbers $|m_{j}| \leq 5/2$. In the following, we will examine quasi-forbidden resonances from initial states with ${m_{j}\!=\!1/2}$ or $3/2$ while the selection rule $\Delta m_{j}=0, \pm 1$ is still preserved for dipole-dipole transitions.\\

In Fig.~\ref{Fig:28p32m32ThGlobal}, we plot a graphical solution of the resonance condition occurring in a F\"orster resonance (see Eqs.~(\ref{eq:ResonanceCondition}) and~(\ref{eq:ForsterResonance})). We transform the resonance condition ${2 \times E_{r_2} = E_{r_1} + E_{r_3}}$ in ${E_{r_1} = 2 \times E_{r_2} - E_{r_3}}$ and for the initial state $\left| r_2 \right> = \left| 28 p_{3/2} m_{j}\!=\!3/2 \right>$ we plot in dashed red the energy corresponding to ${2 \times E_{r_2} - E_{r_3}}$. The resonance condition is then fulfilled at any energy crossing of solid black and dashed red ``states". The resulting plot will be denoted as a resonance map. The advantage of this graphical solution compared to the usual pair state energy plots \cite{2005PRLGallagher} lies in the simplicity to locate the quasi-forbidden resonances in the vicinity of the multiplicity. Indeed when the energies of $N$ states are calculated, for a pair state energy plot \textit{a priori} $N^2$ curves are necessary to verify all the possible resonances, while in our graphical method only $2 N$ curves are required. The basic idea of the usual pair state energy plot is that only a small number of resonances are present and only the authorized ones are plotted. We also acknowledge the work in \cite{2009ThesisvanDitzhuijzen} where a similar plot was used as we discovered during this paper redaction.\\

\begin{figure*}[h]
    \begin{center}
    	\includegraphics[width=12.6 cm]{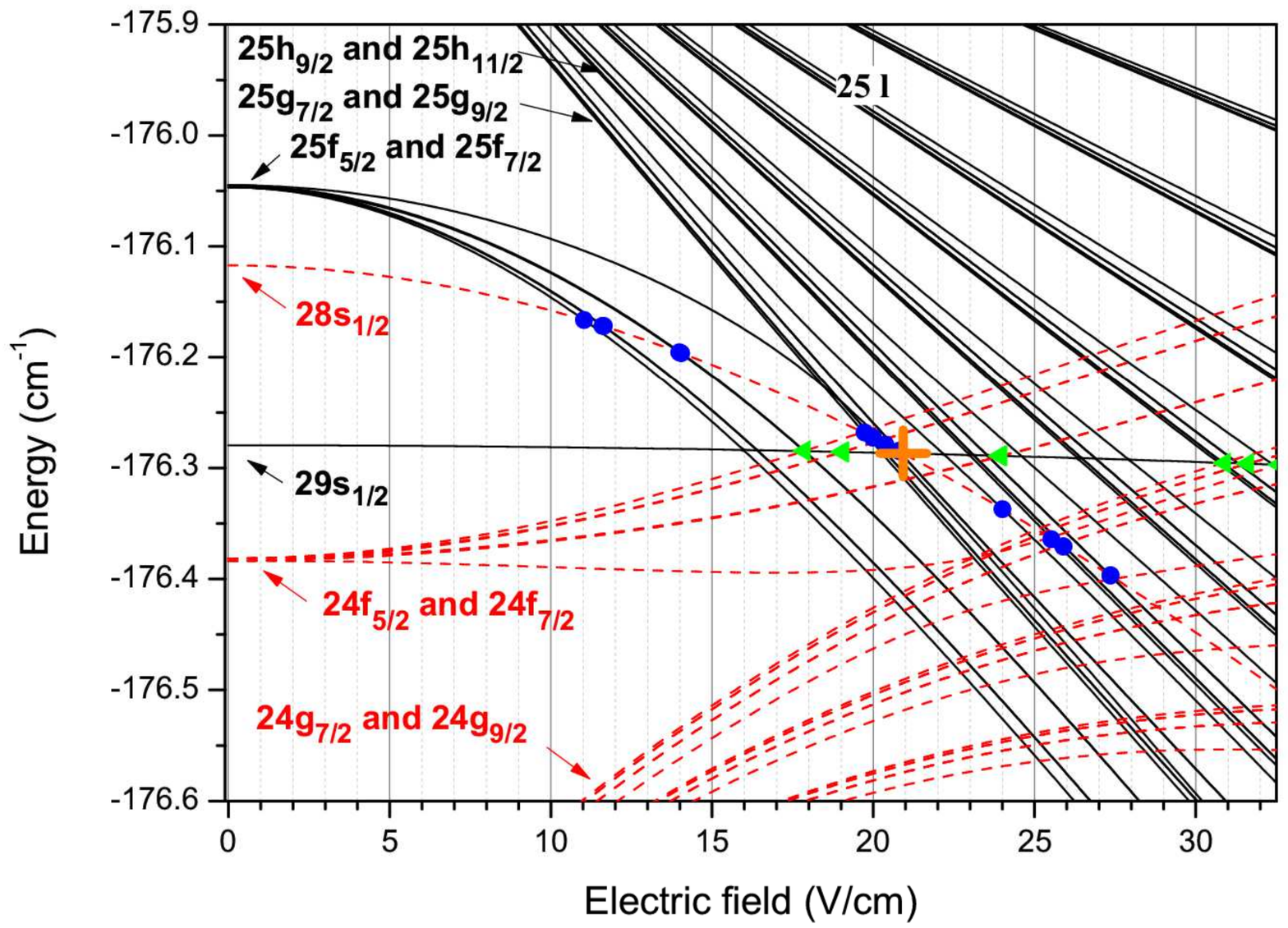}
		\end{center}
		\setlength\abovecaptionskip{-1 cm}
   	\caption{(Color online) Resonance map for the $\left| r_2 \right> = \left| 28 p_{3/2} m_{j}\!=\!3/2 \right>$ initial state in the vicinity of the $29s$ and $28s$ states. In solid black are plotted the eigenenergies $E_{r_1}$ \textit{vs} the applied external electric field $F$. In dashed red, we plot the energy corresponding to ${2 \times E_{r_2} - E_{r_3}}$ to obtain a graphical solution of the resonance condition occurring in a F\"orster resonance. Here the thick orange cross indicates the allowed F\"orster resonance, the light green triangles and the blue circles show the multiple quasi-forbidden F\"orster resonances transferring atoms to the $29s$ and $28s$ states respectively. Only the absolute values of $|m_{j}|$ are plotted.}
    \label{Fig:28p32m32ThForsterResonances28s-25f28s-25g}
\end{figure*}

\begin{figure*}[h]
    \begin{center}
    	\includegraphics[width=11.4 cm]{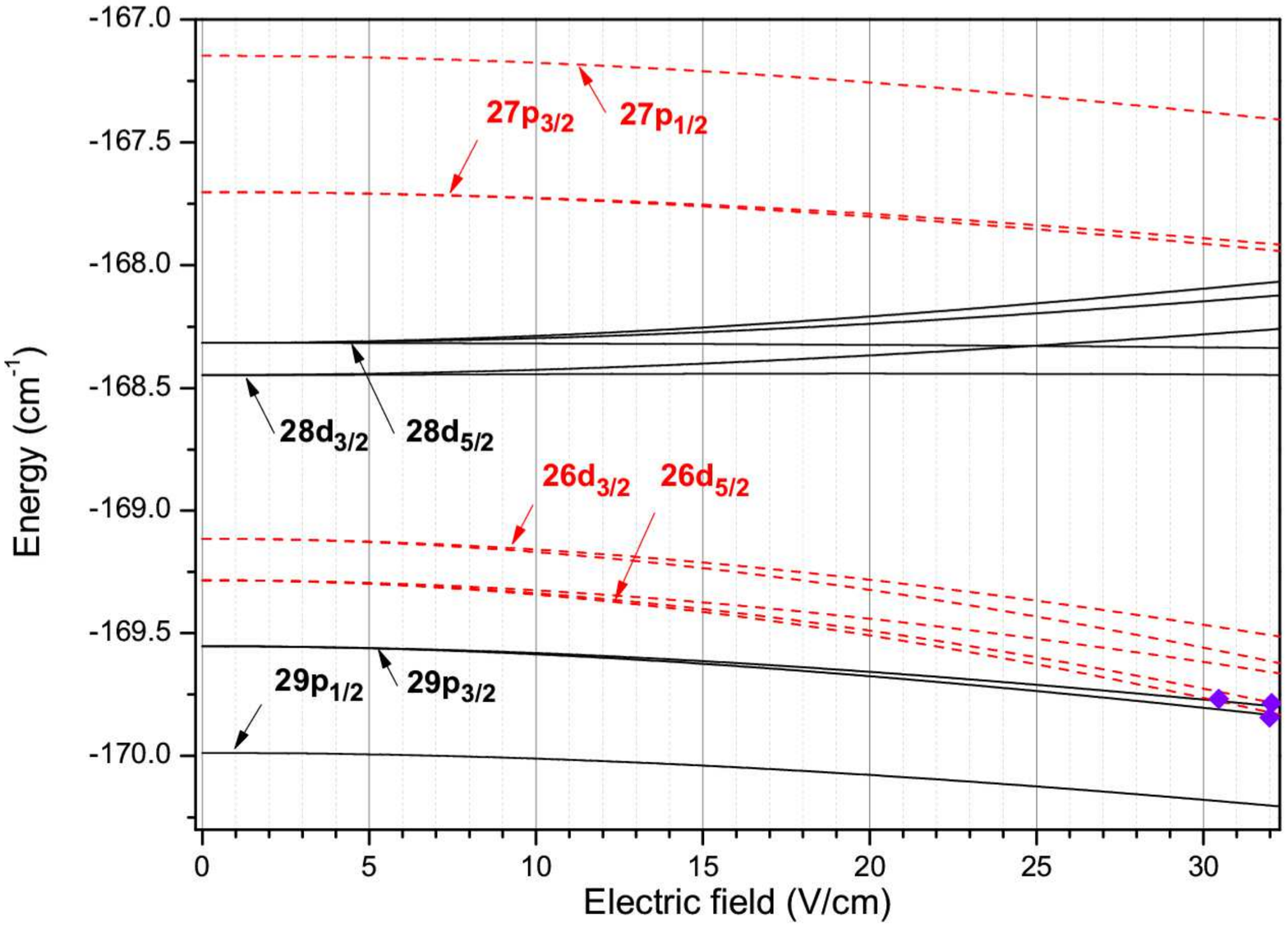}
		\end{center}
		\setlength\abovecaptionskip{-0.5 cm}
   	\caption{(Color online) Resonance map for the $\left| r_2 \right> = \left| 28 p_{3/2} m_{j}\!=\!3/2 \right>$ initial state in the vicinity of the $28d$ and $26d$ states. In solid black are plotted the eigenenergies $E_{r_1}$ \textit{vs} the applied external electric field $F$. In dashed red, we plot the energy corresponding to ${2 \times E_{r_2} - E_{r_3}}$ to obtain a graphical solution of the resonance condition occurring in a F\"orster resonance. Here the purple diamonds show the few quasi-forbidden F\"orster resonances transferring atoms to the $29p$ and $26d$ states. Only the absolute values of $|m_{j}|$ are plotted.}
    \label{Fig:28p32m32ThForsterResonances29p-26d}
\end{figure*}

At moderate electric fields, the strongest quasi-forbidden resonances appear when the initial state and one of the final one fulfil the ${\Delta l = \pm 1}$ selection rule. Therefore we will focus on resonances from the initial $p$ state with at least one $s$ or $d$ final state as shown with light green triangles and blue circles on Fig.~\ref{Fig:28p32m32ThForsterResonances28s-25f28s-25g} and with purple diamonds on Fig.~\ref{Fig:28p32m32ThForsterResonances29p-26d}. Those symbols denote the multiple quasi-forbidden F\"orster resonances, while a thick orange cross indicates an allowed F\"orster resonance.\\



To get the Stark field of the allowed F\"orster resonance starting from the $28 p_{3/2} m_{j}\!=\!3/2$ initial state, we follow on Fig.~\ref{Fig:28p32m32ThForsterResonances28s-25f28s-25g} the dashed red curve labelled ${28 s_{1/2} m_{j}\!=\!1/2}$ (corresponding to the energy of ${2 \times E \left( 28 p_{3/2} m_{j}\!=\!3/2 \right) - E \left( 28 s_{1/2} m_{j}\!=\!1/2 \right)}$) until it crosses the solid black curve representing the ${29 s_{1/2} m_{j}\!=\!1/2}$ state at $F\!=\!20.91$~V/cm. In the same way, we can identify the three main groups of quasi-forbidden F\"orster resonances fulfilling the ${\Delta l = \pm 1}$ condition for one of both final states. For the first group of resonances, we follow on Fig.~\ref{Fig:28p32m32ThForsterResonances28s-25f28s-25g} the dashed red curve labelled ${28 s_{1/2} m_{j}\!=\!1/2}$ until it crosses the ${25 f_{7/2}}$ manifold in solid black at ${F\!=\!11.04; 11.60; 14.00}$~V/cm for the ${m_{j}\!=\!1/2; 3/2; 5/2}$ states respectively. Here only states with a ${\Delta m_{j}=0, \pm 1}$ will be coupled in a dipole-dipole transition, eliminating the transition to the ${m_{j}\!=\!7/2}$ state. In the same group of resonances, the dashed red curve labelled ${28 s_{1/2} m_{j}\!=\!1/2}$ crosses also the ${25 f_{5/2}}$ manifold in solid black at ${F\!=\!11.64; 14.04; 20.38}$~V/cm for the ${m_{j}\!=\!1/2; 3/2; 5/2}$ states respectively. If we continue along the dashed red curve labelled ${28 s_{1/2} m_{j}\!=\!1/2}$, we encounter quasi-forbidden F\"orster resonances crossing the solid black ${25 g_{9/2}}$, ${25 g_{7/2}}$, ${25 h_{11/2}}$, and ${25 h_{9/2}}$ manifolds around ${F \sim 20-21; 20-24; 25-27; 26-27}$~V/cm respectively. For the second group of resonances, we follow on Fig.~\ref{Fig:28p32m32ThForsterResonances28s-25f28s-25g} the solid black curve labelled ${29 s_{1/2} m_{j}\!=\!1/2}$ until it crosses the ${24 f_{7/2}}$ manifold in dashed red at ${F\!=\!17.85; 19.03; 23.97}$~V/cm for the ${m_{j}\!=\!1/2; 3/2; 5/2}$ states respectively, and the ${24 f_{5/2}}$ manifold in dashed red at ${F\!=\!19.06; 24.01; 32.58}$~V/cm for the ${m_{j}\!=\!1/2; 3/2; 5/2}$ states respectively. Then the solid black ${29 s_{1/2} m_{j}\!=\!1/2}$ curve also crosses the dashed red ${24 g_{9/2}}$ and ${24 g_{7/2}}$ manifolds around ${F \sim 31; 31.5}$~V/cm respectively. For the third group of resonances, we follow on Fig.~\ref{Fig:28p32m32ThForsterResonances29p-26d} the dashed red curve labelled ${26 d_{5/2} m_{j}\!=\!1/2}$ (the lowest state in the ${26 d_{5/2}}$ manifold) until it crosses the ${29 p_{3/2}}$ states in solid black at $F\!=\!30.42; 31.99$~V/cm for $m_{j}\!=\!3/2; 1/2$ respectively. Similarly the labelled ${26 d_{5/2} m_{j}\!=\!3/2}$ dashed red curve crosses the ${29 p_{3/2} m_{j}\!=\!3/2}$ state at $F\!=\!32.05$~V/cm.\\


To sum-up, the first quasi-forbidden F\"orster resonances of the groups detailed previously lead for instance to the population transfers:
\begin{eqnarray*}
& & 2 \times 28 p_{3/2} m_{j}\!=\!3/2 \leftrightarrow 28 s_{1/2} m_{j}\!=\!1/2 + 25 f_{7/2} m_{j}\!=\!1/2 \\
& & 2 \times 28 p_{3/2} m_{j}\!=\!3/2 \leftrightarrow 29 p_{3/2} m_{j}\!=\!3/2 + 26 d_{5/2} m_{j}\!=\!1/2.
\end{eqnarray*}
In those resonance equations, shown as an example, we see a transfer from a ``labelled" $p$-state to a $f$-state (though transfers to higher orbital quantum number are possible) or with a transfer involving no change of the $l$-state between the initial and final states. Those resonances correspond to quasi-forbidden F\"orster resonances which do not fulfil the ${\Delta l = \pm 1}$ selection rule in presence of a low electric field.\\

We also calculate the strength of the dipole-dipole coupling for all the different quasi-forbidden resonances found with the graphical resolution. The calculation is realized in the frozen gas regime (fixed atoms)~\cite{1998PRLPillet}, within the Born-Oppenheimer approximation, and considering a possible fixed angle $\theta$ between the electric field axis and the interatomic axis. The total angular momentum of the system must be conserved but as the molecular rotational angular momentum can change, the sum of the $m_{j}$ of the two atoms is not necessarily conserved~\cite{2014PRLMerkt}. We can thus compute the dipole-dipole interaction for all the observed resonances. We then average the dipole-dipole coupling strength over this angle $\theta$ and over the possible signs of all initial $m_{j}$ while we sum the contribution of the possible signs of all final $m_{j}$, as they are degenerate in the electric field and the corresponding resonances will perfectly overlap. To compute this average, we assume a weak coupling regime for which the observed transfer is proportional to the square of the interaction strength. Tables~\ref{tab:QuasiForbiddenResonances28p32m12} and~\ref{tab:QuasiForbiddenResonances28p32m32} summarize the coupling for resonances starting from the $28 p_{3/2} m_{j}\!=\!1/2$ state and $28 p_{3/2} m_{j}\!=\!3/2$ state respectively. Tables~\ref{tab:QuasiForbiddenResonances32p32m12} and~\ref{tab:QuasiForbiddenResonances32p32m32} display the same calculations for $n=32$.

\begin{table*}
\caption{Summary of dipole-dipole coupling strength for quasi-forbidden resonances starting from the $28 p_{3/2} m_{j}\!=\!1/2$ state.\label{tab:QuasiForbiddenResonances28p32m12}}
\begin{ruledtabular}
\begin{tabular}{ccccccccc}
Final states & \multicolumn{2}{c}{$28 s + 25 f_{7/2} m_{j}^{\prime}$} & \multicolumn{2}{c}{$28 s + 25 f_{5/2} m_{j}^{\prime}$} & \multicolumn{2}{c}{$29 s + 24 f_{7/2} m_{j}^{\prime}$} & \multicolumn{2}{c}{$29 s + 24 f_{5/2} m_{j}^{\prime}$} \\
$m_{j}^{\prime}$ & 1/2 & 3/2 & 1/2 & 3/2 & 1/2 & 3/2 & 1/2 & 3/2 \\
Coupling at $1$~$\mu$m (MHz) & 12.25 & 8.95 & 10.4 & 7.7 & 15.5 & 9.98 & 12.45 & 7.05 \\
Resonance (V/cm) & 11.95 & 12.68 & 12.72 & 15.86 & 20.92 & 22.75 & 22.77 & 30.35 \\
\hline
Final states & \multicolumn{2}{c}{$28 s + 25 g_{9/2} m_{j}^{\prime}$} & \multicolumn{2}{c}{$28 s + 25 g_{7/2} m_{j}^{\prime}$} & \multicolumn{3}{c}{$29 p_{3/2} m_{j}^{\prime} + 26 d_{5/2} m_{j}^{\prime\prime}$} & \\
$m_{j}^{\prime}$ or ($m_{j}^{\prime}$, $m_{j}^{\prime\prime}$) & 1/2 & 3/2 & 1/2 & 3/2 & (3/2, 1/2) & (1/2, 1/2) & (3/2, 3/2) & \\
Coupling at $1$~$\mu$m (MHz) & 6.93 & 4.65 & 5.93 & 4.8 & 1.2 & 6.78 & 1.05 &  \\
Resonance (V/cm) & 21.55 & 21.9 & 21.9 & 23.18 & 27.37 & 29.09 & 29.34 & 
\end{tabular}
\end{ruledtabular}
\end{table*}

\begin{table*}
\caption{Summary of dipole-dipole coupling strength for quasi-forbidden resonances starting from the $28 p_{3/2} m_{j}\!=\!3/2$ state.\label{tab:QuasiForbiddenResonances28p32m32}}
\begin{ruledtabular}
\begin{tabular}{ccccccccrc}
Final states & \multicolumn{3}{c}{$28 s + 25 f_{7/2} m_{j}^{\prime}$} & \multicolumn{3}{c}{$28 s + 25 f_{5/2} m_{j}^{\prime}$} & \multicolumn{3}{c}{$28 s + 25 g_{9/2} m_{j}^{\prime}$~~~~~~~~~~~} \\
$m_{j}^{\prime}$ & 1/2 & 3/2 & 5/2 & 1/2 & 3/2 & 5/2 & 1/2 & 3/2 & 5/2 \\
Coupling at $1$~$\mu$m (MHz) & 6.65 & 10.3 & 14.05 & 0.81 & 0.7 & 0.55 & 5.15 & 5.15 & 7.35 \\
Resonance (V/cm) & 11.04 & 11.60 & 14.0 & 11.64 & 14.04 & 20.38 & 19.75 & 19.98 & 20.80 \\
\hline
Final states & \multicolumn{3}{c}{$28 s + 25 g_{7/2} m_{j}^{\prime}$} & \multicolumn{3}{c}{$28 s + 25 h_{11/2} m_{j}^{\prime}$} & \multicolumn{2}{c}{$28 s + 25 h_{9/2} m_{j}^{\prime}$} & \multicolumn{1}{c}{$29 p_{3/2} m_{j}^{\prime} + 26 d_{5/2} m_{j}^{\prime\prime}$} \\
$m_{j}^{\prime}$ or ($m_{j}^{\prime}$, $m_{j}^{\prime\prime}$) & 1/2 & 3/2 & 5/2 & 1/2 & 3/2 & 5/2 & 1/2 & 3/2 & (3/2, 1/2) \\
Coupling at $1$~$\mu$m (MHz) & 0.11 & 0.29 & 0.7 & 3.05 & 6.45 & 8.95 & 0.17 & 0.33 & 2.6 \\
Resonance (V/cm) & 19.98 & 20.8 & 24.0 & 25.54 & 25.93 & 27.35 & 25.93 & 27.35 & 30.42 \\
\hline
Final states & \multicolumn{3}{c}{$29 s + 24 f_{7/2} m_{j}^{\prime}$} & \multicolumn{3}{c}{$29 s + 24 f_{5/2} m_{j}^{\prime}$} & \multicolumn{2}{c}{$29 s + 24 g_{9/2} m_{j}^{\prime}$} & \multicolumn{1}{c}{$29 s + 24 g_{7/2} m_{j}^{\prime}$} \\
$m_{j}^{\prime}$ & 1/2 & 3/2 & 5/2 & 1/2 & 3/2 & 5/2 & 1/2 & 3/2 & 1/2  \\
Coupling at $1$~$\mu$m (MHz) & 9.0 & 12.45 & 15.05 & 0.64 & 0.59 & 0.26 & 3.9 & 7.25 & 0.17 \\
Resonance (V/cm) & 17.85 & 19.03 & 23.97 & 19.06 & 24.01 & 32.58 & 30.86 & 31.62 & 31.62
\end{tabular}
\end{ruledtabular}
\end{table*}

\begin{table*}
\caption{Summary of dipole-dipole coupling strength for quasi-forbidden resonances starting from the $32 p_{3/2} m_{j}\!=\!1/2$ state.\label{tab:QuasiForbiddenResonances32p32m12}}
\begin{ruledtabular}
\begin{tabular}{ccccccccc}
Final states & \multicolumn{2}{c}{$32 s + 29 f_{7/2} m_{j}^{\prime}$} & \multicolumn{2}{c}{$32 s + 29 f_{5/2} m_{j}^{\prime}$} & \multicolumn{4}{c}{$33 p_{3/2} m_{j}^{\prime} + 30 d_{5/2} m_{j}^{\prime\prime}$} \\
$m_{j}^{\prime}$ or ($m_{j}^{\prime}$, $m_{j}^{\prime\prime}$) & 1/2 & 3/2 & 1/2 & 3/2 & (3/2, 1/2) & (1/2, 1/2) & (3/2, 3/2) & (1/2, 3/2) \\
Coupling at $1$~$\mu$m (MHz) & 27.1 & 18.05 & 22.45 & 13.35 & 0.98 & 6.0 & 0.9 & 4.83 \\
Resonance (V/cm) & 8.2 & 8.7 & 8.7 & 10.4 & 4.45 & 4.78 & 4.83 & 5.22
\end{tabular}
\end{ruledtabular}
\end{table*}

\begin{table*}
\caption{Summary of dipole-dipole coupling strength for quasi-forbidden resonances starting from the $32 p_{3/2} m_{j}\!=\!3/2$ state.\label{tab:QuasiForbiddenResonances32p32m32}}
\begin{ruledtabular}
\begin{tabular}{ccccccc}
Final states & \multicolumn{3}{c}{$32 s + 29 f_{7/2} m_{j}^{\prime}$} & \multicolumn{2}{c}{$32 s + 29 f_{5/2} m_{j}^{\prime}$} & \\
$m_{j}^{\prime}$ & 1/2 & 3/2 & 5/2 & 1/2 & 3/2 &  \\
Coupling at $1$~$\mu$m (MHz) & 29.0 & 22.5 & 26.8 & 0.8 & 1.03 & \\
Resonance (V/cm) & 7.5 (or 7.44)\cite{QuantumDefect} & 7.86 & 9.25 & 7.87 & 9.26 &  \\
\hline
Final states & \multicolumn{6}{c}{$33 p_{3/2} m_{j}^{\prime} + 30 d_{5/2} m_{j}^{\prime\prime}$} \\
($m_{j}^{\prime}$, $m_{j}^{\prime\prime}$) & (3/2, 1/2) & (1/2, 1/2) & (3/2, 3/2) & (1/2, 3/2) & (3/2, 5/2) & (1/2, 5/2) \\
Coupling at $1$~$\mu$m (MHz) & 1.73 & 0.05 & 4.18 & 0.08 & 7.55 & 0.16 \\
Resonance (V/cm) & 4.98 & 5.4 & 5.5 & 6.08 & 7.55 & 9.15
\end{tabular}
\end{ruledtabular}
\end{table*}

\clearpage

\section{Experimental set-up}

\begin{figure}[h]
    \begin{center}
    	\includegraphics[width=7.0 cm]{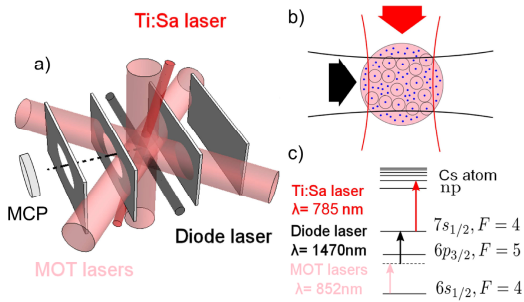}
		\end{center}
		\setlength\abovecaptionskip{-1 cm}
   	\caption{(Color online) a. Experimental set-up with a MOT in the center of four metallic wire mesh grids used to apply electric fields with voltages up to ${\pm 5}$~kV. Rydberg atoms are excited at the cross-section of the 3-photon excitation lasers. The black dashed line defines the trajectory taken by the ionized Rydberg atoms toward the MCP detector. b. Zoom on the excitation region at the laser cross-section. c. $^{133}$Cs energy levels used for the Rydberg excitation.}
    \label{Fig:ExperimentalSetup}
\end{figure}

We realize a magneto-optical trap (MOT) which is located at the center of four parallel 60~mm by 130~mm wire mesh grids of 80~$\mu$m thickness and 1~mm grid spacing (see Fig.~\ref{Fig:ExperimentalSetup} a)) \cite{2012PRLPillet}. In this MOT, cesium atoms are cooled down to $100$~$\mu$K and can be considered as frozen compared to the lifetime of the Rydberg state ($\tau_{\mathrm{rad}} \sim 60$~$\mu$s for $30 p_{3/2}$ and $\tau_{\mathrm{rad}} \sim 21$~$\mu$s for $30 s_{1/2}$ \cite{2009PRABeterov}). The inner pair of grids is spaced by ${1.88 \pm 0.02}$~cm while the outer grids are 1.5~cm apart from the inner grids. At the beginning of the experiment sequence, a small electric field of several V/cm is applied on the inner grids.

\begin{figure}[h]
    \begin{center}
    	\includegraphics[width=7.6 cm]{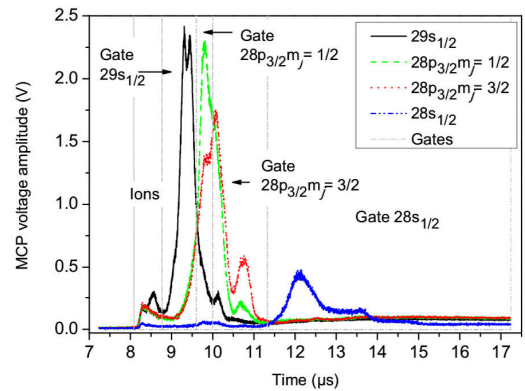}
		\end{center}
		\setlength\abovecaptionskip{-0.4 cm}
   	\caption{(Color online) Time-of-flight signal from selective field ionization (at $V_{\mathrm{ioniz}}=2630$~V) to detect the $np$, $ns$ and $(n+1)s$ signals on the MCP where each state has been directly excited at an electric field far from any F\"orster resonances.} 
    \label{Fig:ToF}
\end{figure}

We excite the atoms to Stark-shifted Rydberg states, $nl$, using a 3-photon transition (see Fig.~\ref{Fig:ExperimentalSetup}~c)). We start from the MOT lasers and use two additional lasers to couple the states: $6s \rightarrow 6p \rightarrow 7s \rightarrow nl$. The $6p \rightarrow 7s$ step uses 10~mW of a 1470 nm diode laser. A cw Ti:sapphire ring laser, providing roughly 600~mW on the atoms at 785~nm and locked on an ultra-stable cavity, drives the $7s \rightarrow nl$ transition. Those two lasers are focused to 300~$\mu$m and 200~$\mu$m spot diameters respectively and are perpendicularly overlapped in the atomic sample (see Fig.~\ref{Fig:ExperimentalSetup} b)). Those beams are switched on for ${\tau_{\mathrm{imp}} \sim 200}$~ns, at a 10 Hz repetition rate, by two different acousto-optic modulators. We thus excite up to $10^{5}$ atoms in the $nl$ state within a $200$~$\mu$m diameter cloud having a typical density of $10^{10}$~cm$^{-3}$. We let the atoms interact \textit{via} dipole-dipole interaction during a delay of ${\tau_{\mathrm{delay}} \sim 1}$~$\mu$s after the Rydberg excitation and realize a selective field ionization (SFI) of the Rydberg atoms. We choose a short enough delay to avoid resonance broadening due to electric field inhomogeneities induced by ions created by Penning ionization and blackbody radiations. A high voltage ramp is then applied on the inner back grid, rising to ${V_{\mathrm{ioniz}}=2.6; 1.35}$~kV for ${n=28; 32}$ in $4$~$\mu$s and ionizing the various Rydberg levels at different times. After a flight of 210~mm from the center of the trapped cloud, ions are detected by a micro-channel plate (MCP) detector. The amplitude of the field ionization pulse is chosen to optimally isolate the $np$ time-of-flight (TOF) signal from the other $ns$ and $(n+1)s$ signals (from Eq.~(\ref{eq:ForsterResonance})). A typical TOF for all involved states in the allowed F\"orster resonance at $n\!=\!28$ is displayed on Fig.~\ref{Fig:ToF} where each state has been excited independently at an electric field far from any F\"orster resonance to avoid any transfer. We use these TOF references to define temporal gates corresponding to each state and compute the cross-talks between the different gates. During a measurement, we only excite the atoms to the $n p_{3/2}$ states and extract the state population from their temporal gates to determine the fraction of total population in different channels (corrected from their cross-talks): $ns$, $n p_{3/2} m_{j}\!=\!1/2$, $n p_{3/2} m_{j}\!=\!3/2$, and $(n+1)s$. This analysis gives very accurate results on the population transfer for dipole allowed F\"orster resonances \cite{2012PRLPillet}. For quasi-forbidden resonances, other states are created and this analysis is no longer valid. An additional channel, the ion gate, was initially used to quantify the number of ions created before the field ionization, but it is used here to show atoms that are in an upper level than the ${(n+1)s}$. When atoms with higher energy states are created, they add up in the ion gate. Their transfer is then generally underestimated. To get a more precise transfer efficiency, we would need to record a specific set of TOF references for each resonance. Nonetheless using our state population analysis, we can identify the resonances and estimate their relative amplitudes.


\section{Quasi-forbidden F\"orster resonances}

Tuning the static electric field, 2-body F\"orster resonances can be scanned due to the Stark shift of Rydberg levels~\cite{1983PRAGallagher, 1994PRAGallagher}. Those resonances are then detected by SFI since they produce changes in the population of the different states monitored on the TOF signals.\\

\begin{figure*}[h]
    \begin{center}
    	\includegraphics[width=15.0 cm]{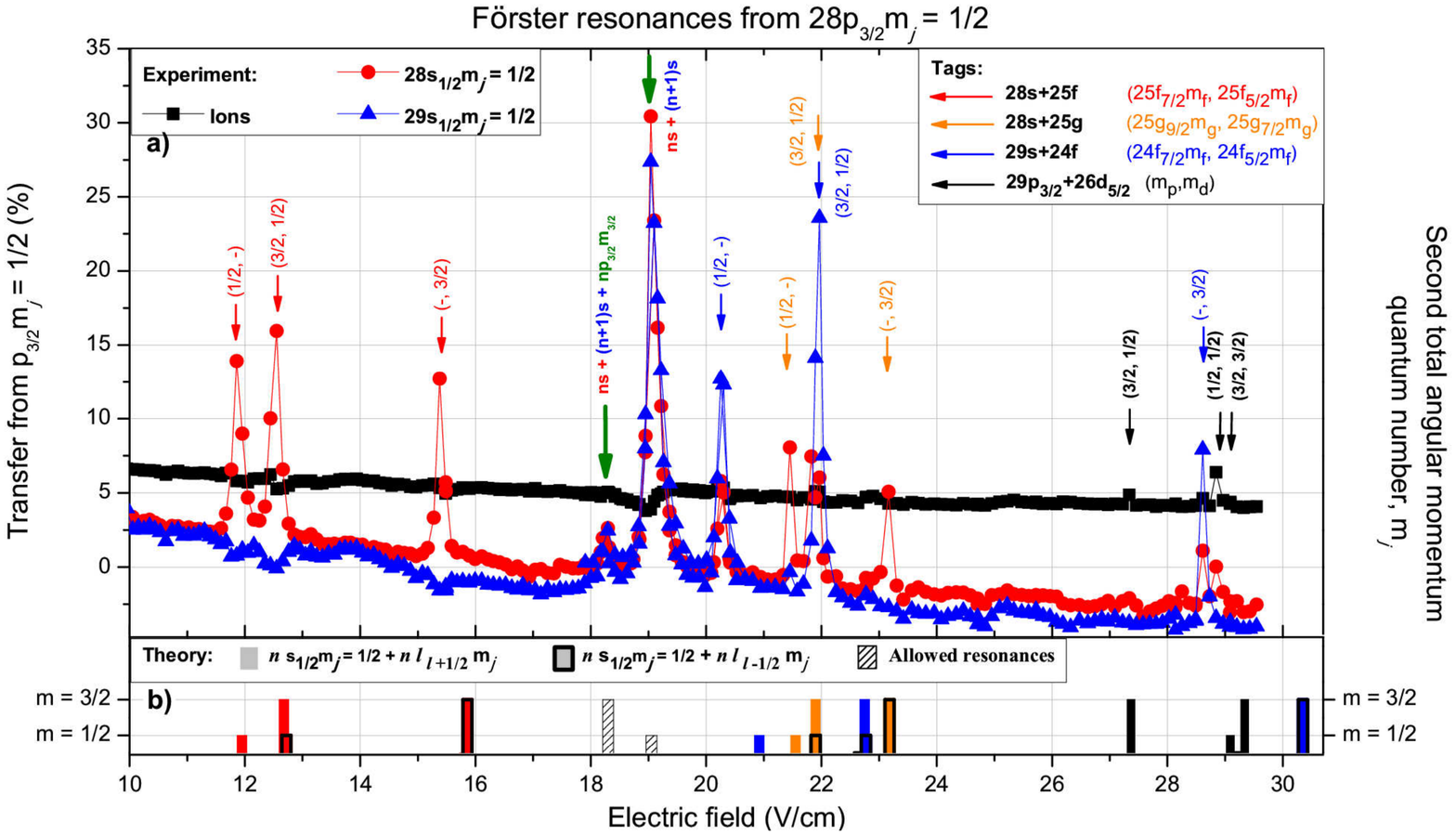}
		\setlength\abovecaptionskip{-1.5 cm}
		\setlength\belowcaptionskip{-2.0 cm}
   	\caption{(Color online) Quasi-forbidden F\"orster resonances around the allowed F\"orster resonance ${2 \times 28 p_{3/2} m_{j}\!=\!1/2 \leftrightarrow 28 s_{1/2} m_{j}\!=\!1/2 + 29 s_{1/2} m_{j}\!=\!1/2}$ located at $F\!=\!19.05$~V/cm. a) Experimental measurement where all final states of the different resonances are tagged. Ion gate is not included in the cross-talk correction so the mean baseline is not accurate. b) Bar diagram representing the theoretical electric field resonance positions. We emphasize that the bar amplitude codes the projection $m_{j}$ of the total angular momentum of the final state having $J > 1/2$. Moreover we distinguish the $l_{l+1/2}$ and $l_{l-1/2}$ final states adding a frame.}
    \label{Fig:28p32m12}
    \end{center}
\end{figure*}

We plot in Fig.~\ref{Fig:28p32m12} a) the relative population of atoms transferred into various gates: the ${28 s_{1/2} m_{j}\!=\!1/2}$ gate (in red), the ${29 s_{1/2} m_{j}\!=\!1/2}$ gate (in blue), and in the ion gate (in black) depending on the applied electric field, $F$, from atoms initially prepared in the ${28 p_{3/2} m_{j}\!=\!1/2}$ state. In order to determine the baselines, we take a reference measurement at an electric field where no resonance is expected, here at $F\!=\!19.8$~V/cm. On this graph, the allowed F\"orster resonance described below is expected and observed at $F\!=\!19.05$~V/cm:
\begin{equation*}
2 \times 28 p_{3/2} m_{j}\!=\!1/2 \leftrightarrow 28 s_{1/2} m_{j}\!=\!1/2 + 29 s_{1/2} m_{j}\!=\!1/2.
\end{equation*}
Another expected resonance at $F\!=\!18.3$~V/cm is a {3-body} process described in~\cite{2015NatureCommPillet}. All other resonances are quasi-forbidden F\"orster resonances. Some of them concern a transfer to $28 s$ and another state with $\Delta l > 1$:
\begin{eqnarray*}
2 \times 28 p_{3/2} m_{j}\!=\!1/2 & \leftrightarrow & 28 s_{1/2} m_{j}\!=\!1/2 + 25 f_{j} m_{j}^{\prime} \\
2 \times 28 p_{3/2} m_{j}\!=\!1/2 & \leftrightarrow & 28 s_{1/2} m_{j}\!=\!1/2 + 25 g_{j} m_{j}^{\prime}.
\end{eqnarray*}
Other ones correspond to a transfer to $29 s$ and $24 f$: 
\begin{equation*}
2 \times 28 p_{3/2} m_{j}\!=\!1/2 \leftrightarrow 29 s_{1/2} m_{j}\!=\!1/2 + 24 f_{j} m_{j}^{\prime}.
\end{equation*}
For the resonances on Fig.~\ref{Fig:28p32m12}, $f_{j}$ takes the values $f_{7/2}$ and $f_{5/2}$, while $g_{j}$ is $g_{9/2}$ or $g_{7/2}$, and $m_{j}^{\prime}$ is $1/2$ or $3/2$.\\

\begin{figure*}[h]
    \begin{center}
    	\includegraphics[width=15.0 cm]{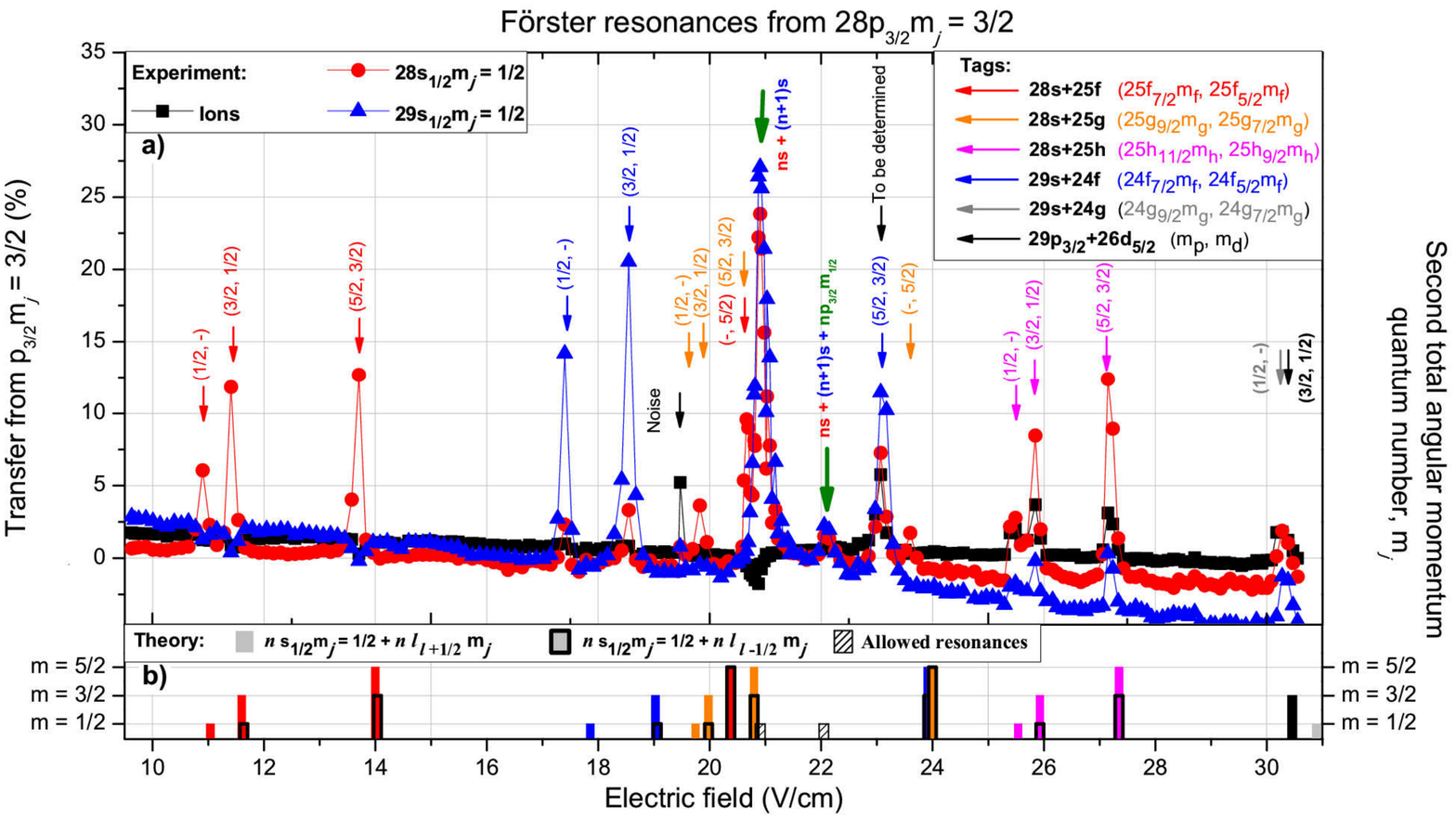}
		\setlength\abovecaptionskip{-1.5 cm}
		\setlength\belowcaptionskip{-0.5 cm}
   	\caption{(Color online) Quasi-forbidden F\"orster resonances around the allowed F\"orster resonance ${2 \times 28 p_{3/2} m_{j}\!=\!3/2 \leftrightarrow 28 s_{1/2} m_{j}\!=\!1/2 + 29 s_{1/2} m_{j}\!=\!1/2}$ located at $F\!=\!20.91$~V/cm. a) Experimental measurement where all final states of the different resonances are tagged. Ion gate is not included in the cross-talk correction so the mean baseline is not accurate. b) Bar diagram representing the theoretical electric field resonance positions. We emphasize that the bar amplitude codes the projection $m_{j}$ of the total angular momentum of the final state having $J > 1/2$. Moreover we distinguish the $l_{l+1/2}$ and $l_{l-1/2}$ final states adding a frame.}
    \label{Fig:28p32m32}
    \end{center}
\end{figure*}

In Fig.~\ref{Fig:28p32m32} a), we realize a similar measurement from atoms initially excited in the ${28 p_{3/2} m_{j}\!=\!3/2}$ state, with a reference at $F\!=\!21.7$~V/cm. We find most of the formerly described resonances shifted in electric field. Indeed the ${28 p_{3/2} m_{j}\!=\!3/2}$ state, higher in energy, requires a higher electric field to be Stark-shifted down to the allowed resonance. On this graph, the following allowed F\"orster resonance is expected and observed at $F\!=\!20.91$~V/cm:
\begin{equation*}
2 \times 28 p_{3/2} m_{j}\!=\!3/2 \leftrightarrow 28 s_{1/2} m_{j}\!=\!1/2 + 29 s_{1/2} m_{j}\!=\!1/2,
\end{equation*}
while the 3-body process is expected at $F\!=\!22.05$~V/cm. Among the formerly described quasi-forbidden F\"orster resonances, an additional group of resonances toward the ``labelled" $h$-states is also present
\begin{equation*}
2 \times 28 p_{3/2} m_{j}\!=\!3/2 \leftrightarrow 28 s_{1/2} m_{j}\!=\!1/2 + 25 h_{j} m_{j}^{\prime}.
\label{eq:ForbiddenForsterResonance28p32m32}
\end{equation*}
For the different resonances on Fig.~\ref{Fig:28p32m32}, $f_{j}$ takes the values $f_{7/2}$ and $f_{5/2}$, while $g_{j}$ is $g_{9/2}$ or $g_{7/2}$, $h_{j}$ is $h_{11/2}$ or $h_{9/2}$, and $m_{j}^{\prime}$ ranges from $1/2$ to $5/2$.\\

All theoretical predictions of those resonances are represented on Fig.~\ref{Fig:28p32m12} b) and Fig.~\ref{Fig:28p32m32} b) as bar diagrams. They define the locations in Stark field of F\"orster resonances thanks to the graphical method presented in Section~\ref{sec:Dip-dip} (see Fig.~\ref{Fig:28p32m32ThForsterResonances28s-25f28s-25g}, \ref{Fig:28p32m32ThForsterResonances29p-26d}, Tables~\ref{tab:QuasiForbiddenResonances28p32m12},~\ref{tab:QuasiForbiddenResonances28p32m32},~\ref{tab:QuasiForbiddenResonances32p32m12} and~\ref{tab:QuasiForbiddenResonances32p32m32}). Due to the large number of close resonances, we choose to code the projection of the total angular momentum quantum number $m_{j}$ in the height of the bars (for the final states having a first total angular momentum quantum number $J > 1/2$), \textit{i.e.} in Fig.~\ref{Fig:28p32m12} the lower level represents $m_{j}\!=\!1/2$ and the higher level $m_{j}\!=\!3/2$, while in Fig.~\ref{Fig:28p32m32} the lower level represents $m_{j}\!=\!1/2$, the intermediate level $m_{j}\!=\!3/2$, and the higher level $m_{j}\!=\!5/2$.\\

From Fig.~\ref{Fig:28p32m12} and Fig.~\ref{Fig:28p32m32}, we can see both allowed 2-body F\"orster resonances which are saturated close to $25\%$ on each state $ns$ and $(n+1)s$, giving a total transfer of $50\%$. This value is the expected saturation due to the random distribution in distances between Rydberg atoms, leading to a statistical average between pairs of atoms in $np$ and pairs in $ns+(n+1)s$. The quasi-forbidden F\"orster resonances appear clearly on this spectrum with a maximum change in the orbital quantum number from ``labelled" $p$-states to $h$-states. They show for a single resonance an estimated non-negligible total transfer up to $30 \%$ with a large uncertainty (see Section~\ref{sec:Discussion}), as only one of the two final states is correctly detected with a transfer of $15 \%$. We also note that while this coupling is mainly due to dipole-dipole interactions allowed by the electric field induced $l$-mixing, the resonances might contain additional dipole-multipole contributions~\cite{2014PRLMerkt} which cannot be distinguished experimentally here. We find those resonances close to the theoretical predictions, with a discrepancy compatible with the uncertainty in the quantum defects and in our field calibration (see Section~\ref{sec:Discussion}). In order to identify some of them, the analysis of the TOF signal shape was necessary to determine the resonance as discussed in Section~\ref{sec:Discussion}.\\

\begin{figure*}[h]
    \begin{center}
    	\includegraphics[width=13.2 cm]{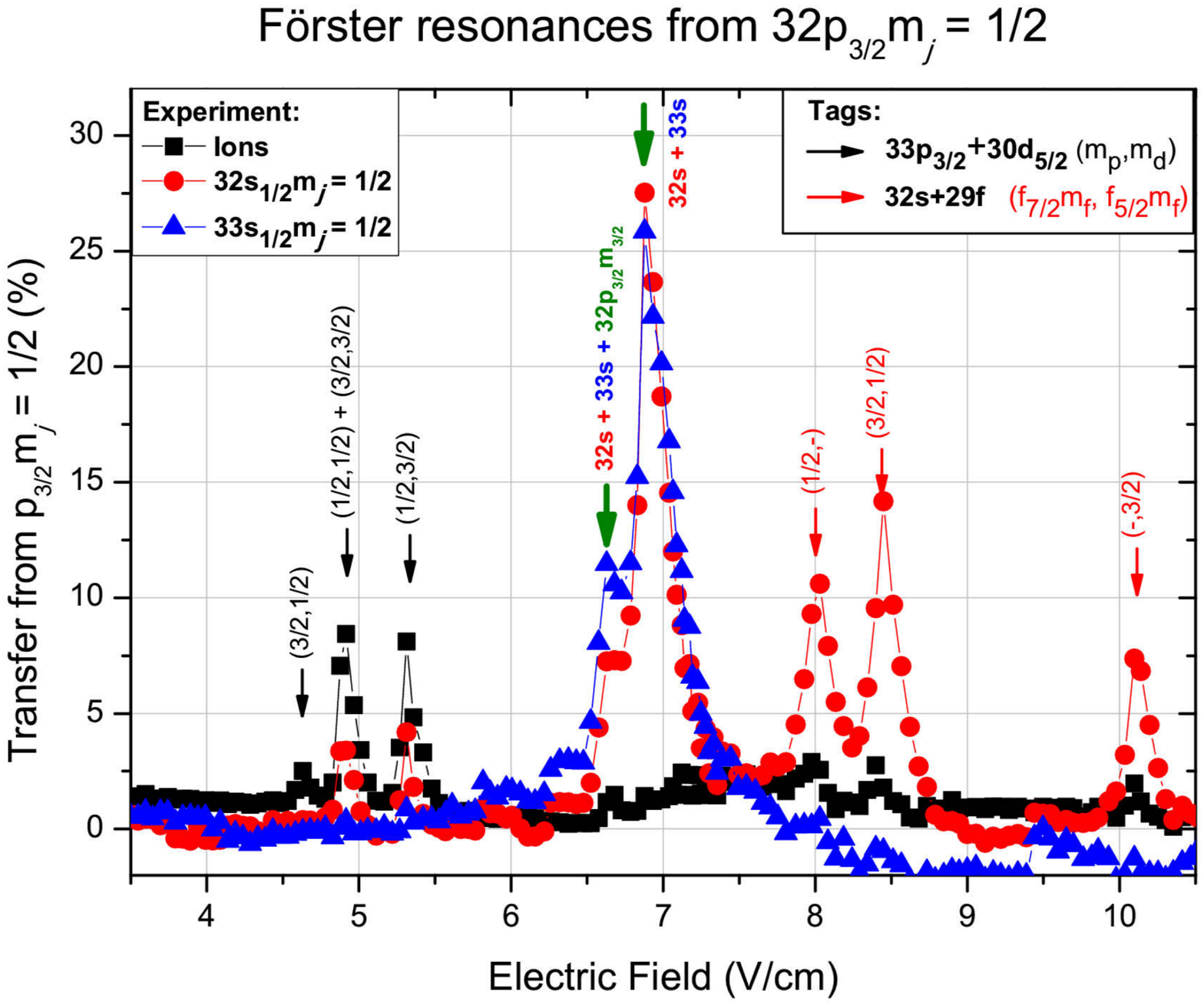}
		\setlength\abovecaptionskip{-0.5 cm}
   	\caption{(Color online) Quasi-forbidden F\"orster resonances around the allowed F\"orster resonance ${2 \times 32 p_{3/2} m_{j}\!=\!1/2 \leftrightarrow 32 s_{1/2} m_{j}\!=\!1/2 + 33 s_{1/2} m_{j}\!=\!1/2}$ located at $F\!=\!6.89$~V/cm. Here is shown the experimental measurement where all final states of the different resonances are tagged. Ion gate is not included in the cross-talk correction so the mean baseline is not accurate.}
    \label{Fig:32p32m12}
    \end{center}
\end{figure*}

\begin{figure*}[h]
    \begin{center}
    	\includegraphics[width=13.2 cm]{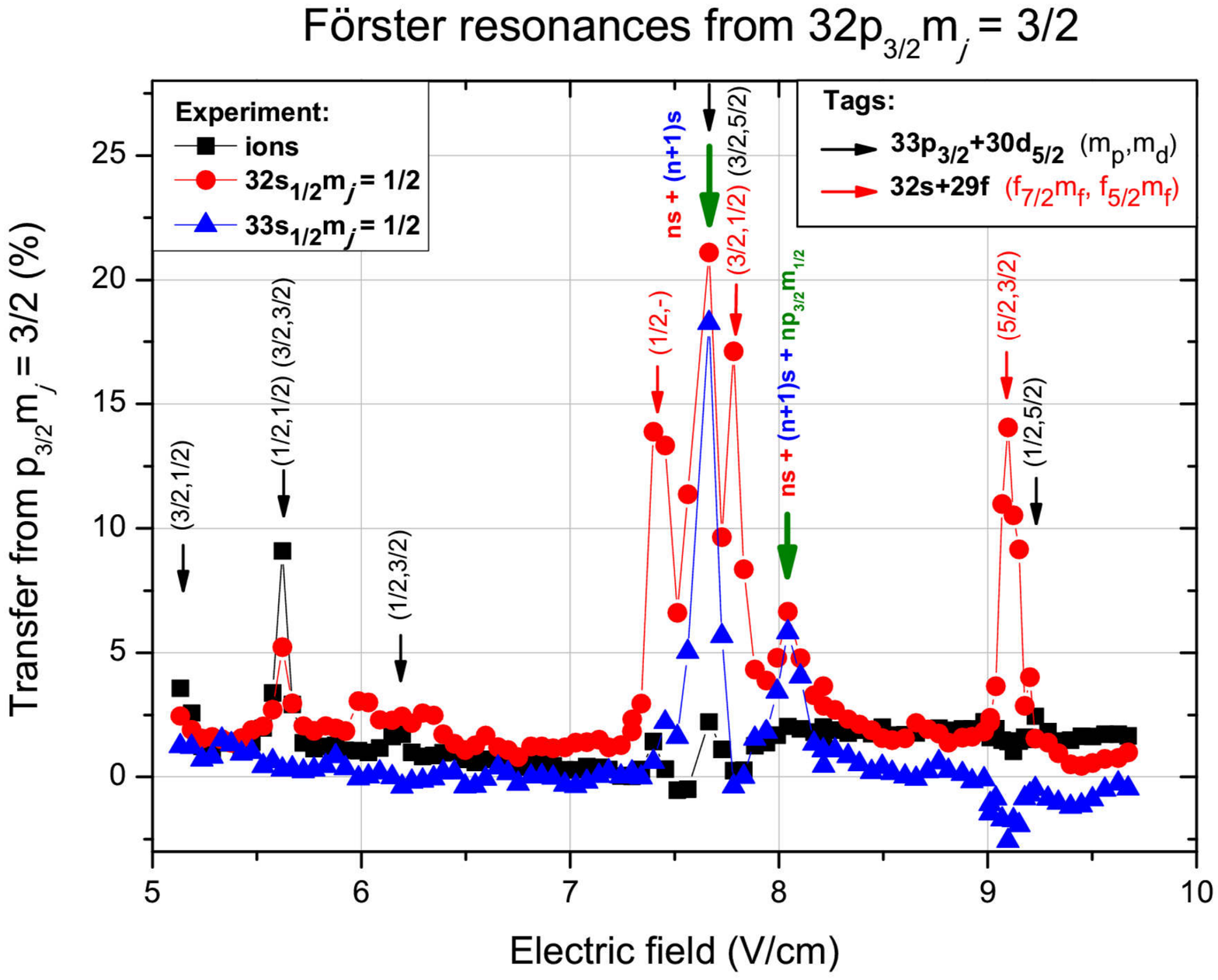}
		\setlength\abovecaptionskip{-0.5 cm}
   	\caption{(Color online) Quasi-forbidden F\"orster resonances around the allowed F\"orster resonance ${2 \times 32 p_{3/2} m_{j}\!=\!3/2 \leftrightarrow 32 s_{1/2} m_{j}\!=\!1/2 + 33 s_{1/2} m_{j}\!=\!1/2}$ located at $F\!=\!7.55$~V/cm. Here is shown the experimental measurement where all final states of the different resonances are tagged. Ion gate is not included in the cross-talk correction so the mean baseline is not accurate.}
    \label{Fig:32p32m32}
    \end{center}
\end{figure*}

Moreover, quasi-forbidden F\"orster resonances are expected in the vicinity of the allowed F\"orster resonance up to $n\!=\!32$, as shown on Fig.~\ref{Fig:32p32m12} and Fig.~\ref{Fig:32p32m32} where references are taken at $F\!=\!5.6$~V/cm and $F\!=\!6.8$~V/cm. For higher $n$, the quasi-forbidden resonances are expected further apart from the allowed resonance. At $n\!=\!32$, the involved allowed F\"orster resonances are described by:
\begin{eqnarray*}
& & 2 \times 32 p_{3/2} m_{j}\!=\!1/2 \leftrightarrow 32 s_{1/2} m_{j}\!=\!1/2 + 33 s_{1/2} m_{j}\!=\!1/2\\
& & 2 \times 32 p_{3/2} m_{j}\!=\!3/2 \leftrightarrow 32 s_{1/2} m_{j}\!=\!1/2 + 33 s_{1/2} m_{j}\!=\!1/2.
\label{eq:ForsterResonance32p32}
\end{eqnarray*}
They are expected and observed at $F\!=\!6.89$~V/cm and $F\!=\!7.55$~V/cm respectively, while forbidden resonances from ``labelled" $p$-states to $f$-states show a resonant coupling up to roughly $8 \%$ in the $32s$ gate. We also see quasi-forbidden resonances similar to those seen on Fig.~\ref{Fig:28p32m32ThForsterResonances29p-26d} which are described by the population transfer:
\begin{equation*}
2 \times 32 p_{3/2} m_{j}\!=\!1/2 \leftrightarrow 33 p_{3/2} m_{j}^{\prime} + 30 d_{5/2} m_{j}^{\prime \prime}
\end{equation*}
where $m_{j}^{\prime}$ and $m_{j}^{\prime \prime}$ take the values $1/2$ and $3/2$ in Fig.~\ref{Fig:32p32m12}, while in Fig.~\ref{Fig:32p32m32} $m_{j}^{\prime \prime}$ ranges from $1/2$ to $5/2$. Those resonances experiencing no change in their $l$-state show also a maximum transfer efficiency of about $8 \%$ in the ion gate, corresponding to atoms transferred in the $d$-state.


\clearpage

\section{Discussion on the interpretation of each resonance}
\label{sec:Discussion}

In order to identify the process involved in each resonance, we examine very carefully the TOF signal in order to identify each of the appearing Rydberg atom populations. In most cases, this information combined with the expected patterns of the resonances enables their interpretation with no ambiguity, although the $(n-3)f$ states mainly overlap with the initial $np$ state in the TOF signal. However, some situations are more complex when they involve none of the expected $s$ state or when the resonances are very close. For instance, we will discuss the case of the quasi-forbidden F\"orster resonance located at $F\!=\!23.1$~V/cm on Fig.~\ref{Fig:28p32m32} where the following energy exchange processes have been identified:
\begin{eqnarray*}
2 \times 28 p_{3/2} m_{j}\!=\!3/2 & \leftrightarrow & 29 s_{1/2} m_{j}\!=\!1/2 + 24 f_{7/2} m_{j}\!=\!5/2 \\
2 \times 28 p_{3/2} m_{j}\!=\!3/2 & \leftrightarrow & 29 s_{1/2} m_{j}\!=\!1/2 + 24 f_{5/2} m_{j}\!=\!3/2.
\label{eq:ForbiddenForsterResonance231Vcm}
\end{eqnarray*}

\begin{figure}[h]
    \begin{center}
    	\includegraphics[width=7.6 cm]{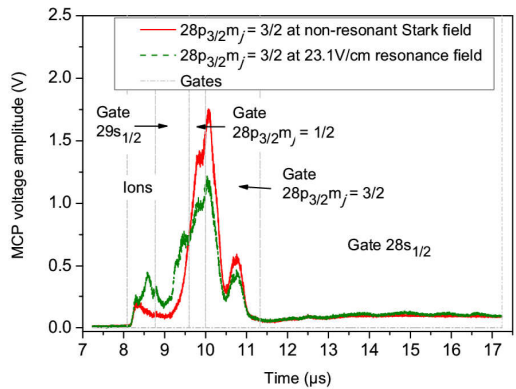}
		\end{center}
		\setlength\abovecaptionskip{-0.4 cm}
   	\caption{(Color online) TOF signal from SFI (at $V_{\mathrm{ioniz}}=2630$~V) to detect the $np$, $ns$ and $(n+1)s$ signals where only the $np$ states have been excited at a quasi-forbidden resonant electric field $F\!=\!23.1$~V/cm (in green). We compare it to the reference signal (in red) shown on Fig.~\ref{Fig:ToF} and see some transfer into the ions, the $(n+1)s$ and $ns$ gates. A lower state seems also to be present in the latter gate.}
    \label{Fig:ToF231Vcm}
\end{figure}

Looking at the TOF signal on this resonance presented in Fig.~\ref{Fig:ToF231Vcm}, we can recognize the main $28p$ state that has been excited (in both red and green traces). On the green trace taken at $F\!=\!23.1$~V/cm, we identify the presence of transferred atoms due to F\"orster resonance: in the $29s$ gate we see the transferred $29s$ atoms, whereas in the $28s$ gate the atoms should correspond to the $24 f$ (being the $(n-4)f$ state) state which ionizes partially in this gate. In addition, we clearly observe a higher state in the ion gate, in comparison with the reference states on Fig.~\ref{Fig:ToF}. As we do not know precisely the TOF of the $24 f$, not observable with our excitation from the $7s$ state, there might also be an even lower energy state than the $24 f$ state in the $28s$ gate in Fig.~\ref{Fig:ToF231Vcm}. It thus seems that a third process occurs at this field which was not anticipated in Fig.~\ref{Fig:28p32m32ThForsterResonances28s-25f28s-25g} and Fig.~\ref{Fig:28p32m32ThForsterResonances29p-26d}. It could then involve states further in energy from the starting $28p$ state like in the following resonance which occurs at the same Stark field:
\begin{equation*}
2 \times 28 p_{3/2} m_{j}\!=\!3/2 \leftrightarrow 27 f_{7/2} m_{j}\!=\!1/2 + 25 d_{5/2} m_{j}\!=\!3/2.
\end{equation*}
However, if we assume this resonance to be strong enough to be observed, we then expect to see other resonances involving all the different $m_{j}$ but we do not observe them.\\

Another peculiar feature appearing on the Fig.~\ref{Fig:28p32m32} is the presence of some signal in the ion gate at $F\!=\!19.4$~V/cm although no resonance is expected. When trying to identify the resonance in a TOF measurement, no transfer was observed. On other measurements not shown here, this signal is not present. Moreover this signal corresponds to a single measurement point while resonances are generally broader. We thus interpret this data as an unexplained noisy measurement.\\

As we see on Fig.~\ref{Fig:28p32m12} and~\ref{Fig:28p32m32}, there is a discrepancy between theory and measurements for some resonances. The two main uncertainties in this comparison between theory and experiment come from the quantum defect uncertainty used for the calculation and the uncertainty of the applied electric field in the experiment. The later is known at $1$ or $2\%$, depending on the electric field, which corresponds to $0.6$~V/cm at maximum in our measurements. Concerning the former, some resonances (to the final states $s+f$ and $s+g$) are systematically observed at a lower electric field than expected which is coherent with the uncertainty of the quantum defects used in the calculation.\\

On Fig.~\ref{Fig:28p32m12},~\ref{Fig:28p32m32},~\ref{Fig:32p32m12}, and~\ref{Fig:32p32m32}, we observe a residual baseline drift probably due to slowly varying ionization path and residual blackbody radiation transfer. One of the limitations of our state population analysis based on a cross-talk estimation lies in the fact that it is realized at one specific electric field (where there is no F\"orster resonance). Then when we change the electric field as plotted in Fig.~\ref{Fig:28p32m12},~\ref{Fig:28p32m32},~\ref{Fig:32p32m12}, and~\ref{Fig:32p32m32}, the cross-talks might be a bit different. The first reason is that we are starting from a different initial voltage for the ionization ramp, which might lead to a different ionization path and then a different TOF shape for each of the state. We try to avoid this problem by setting the voltage to $0$~V before the start of the ionization ramp but there might be a residual voltage. Then the second possible reason concerns the blackbody radiation ionization which might have a different efficiency depending of the applied electric field. Indeed by changing the applied electric field, we change the $l$-mixing and allow then more transitions for the blackbody radiations. This leads to a slightly different state population depending on the applied electric field, which translates to different cross-talks as we change the applied electric field.\\

In the comparison between the observed transfer efficiencies of the quasi-forbidden resonances and their calculated interaction strengths, most features are well reproduced. The small differences should be ascribed to few experimental and theoretical issues. First of all, because the gate analysis is not precisely adapted to each state produced by the various resonances, the measured transfers have an indetermination very hard to evaluate. Indeed for most of the resonances, we only detect correctly half of the final states (the $s$ or the $d$ states) with an uncertainty coming mainly from the imperfect cross-talk compensation between temporal gates. Then we assume that the quasi-forbidden resonances transfer the same atom number in the second final states (the $f, g, h$ or the $p$ states respectively), leading to a quite large total transfer efficiency uncertainty. A good example of this issue is shown in Fig.~\ref{Fig:ToF} where even a $m_j$ state change modifies significantly the TOF output. This issue is particularly important for states as the $(n-3)f$ ones which ionize mainly within the $p$ gates. The second issue playing a role within the calculated strengths is the weak coupling assumption certainly not valid for the strongest resonances. In addition, the average over all signs of initial $m_j$ with equal weights might not be valid as the experiment might generate an imbalance. Finally, we have not considered the role of the small but finite MOT magnetic field, splitting the resonances by at most $10$~mV/cm~\cite{2015PRABrowaeys}. Within our set-up, such a splitting cannot be observed directly and its role on the final transfer efficiency is not at all clear.

\section{Conclusion}

To conclude, we have seen that in the vicinity of the allowed 2-body F\"orster resonance many quasi-forbidden 2-body F\"orster resonances are also present. They show an estimated total transfer up to $30 \%$ (taking into account both final states), with a calculated dipole-dipole coupling around $10\%$ of the dipole allowed resonance coupling. In presence of a low electric field, the dipole-dipole coupling in those quasi-forbidden F\"orster resonances allow population transfer with a change in the orbital quantum number from a ``labelled" $p$-state to a ``labelled" $h$-state or allow transfer involving no change of the $l$-state between the initial and final states. We can assume that at a higher field $\Delta l > 4$ resonances due to $l$-mixing would appear as presented in~\cite{1983PRAGallagher, 1994PRAGallagher}. We have also elaborated a graphical resolution of the resonance condition of F\"orster resonances, allowing to identify clearly the position of the different quasi-forbidden F\"orster resonances. Then we calculated the dipole-dipole coupling strength for all of the observed quasi-forbidden resonances which correspond to their measured relative amplitudes.\\

When using the dipole allowed F\"orster resonances for quantum computation, those quasi-forbidden resonances could perturb the allowed resonance and should be taken into account to determine the total interaction strength in presence of an electric field. Moreover those quasi-forbidden 2-body F\"orster resonances could be of interest in the case of potential processes requiring tunable interactions over a broad band of electric field, like in the search for few-body interactions or to realize macro-molecules built from Rydberg atoms. Indeed, they increase dramatically the number of addressable resonant energy transfers in the cesium atom with an efficiency about $10\%$ of the well-known dipole allowed resonances. For instance, it could increase the number of few-body transfer cascades as demonstrated in~\cite{2012PRLPillet}.

\begin{acknowledgments}

We acknowledge D. Comparat and A. Crubelier for useful discussions on dipole-dipole interactions. E.A. acknowledges the financial support by the Triangle de la Physique (Orsay). This work was supported by the \textit{Agence Nationale de la Recherche} (ANR) under the program COCORYM (ANR-12-BS04-0013) and the EU Marie-Curie program ITN COHERENCE FP7-PEOPLE-2010-ITN.




\end{acknowledgments}


\begin{thebibliography}{38}
\bibitem{QuantumDefect}
At the theoretical resonance field, we observe a strong avoided crossing between the considered $29f$ state and the $33s$ state which will then greatly enhance the calculated interaction. But the observed separation between the quasi-forbidden resonance and the allowed resonance demonstrates that this avoided crossing is further away. In order to compute the correct interaction, we had to modify the $s$ state quantum defect by $10^{-4}$ to obtain the correct relative positions of the resonance (presented in brackets). For consistency with the other values in the table, we also display the original predicted resonance field.
\end{thebibliography}

\end{document}